\documentclass[11pt]{article}
\usepackage{amssymb}
\usepackage{amsmath}

\catcode`@=11
\renewcommand{\section}{\@startsection{section}{1}{0pt}{\medskipamount}
{\medskipamount}{\large\bf}}

\def\de{\delta}
\def\eps{\epsilon}
\def\h{\eta}
\def\th{\vartheta}
\def\la{\lambda}

\def\s{\sigma}
\def\p{\phi}
\def\vp{\varphi}
\def\c{\chi}
\def\j{\psi}

\newcommand{\C}{\mathbb C}
\newcommand{\R}{\mathbb R}

\def\e{\mbox{e}}
\def\im{\mbox{i}}
\def\pa{\mbox{$\partial$}}
\def\diff{\mbox{d}}
\def\sfrac#1#2{{\textstyle\frac{#1}{#2}}}
\def\+{\dagger}

\newcommand{\non}{\nonumber}
\newcommand{\und}{\quad\text{and}\quad}
\newcommand{\with}{\quad\text{with}\quad}
\newcommand{\for}{\quad\text{for}\quad}

\begin{document}

\begin{titlepage}
\setcounter{page}{0}
\begin{flushright}
hep-th/0412042\\
ITP--UH--27/04\\
\end{flushright}

\vskip 2.0cm

\begin{center}

{\Large\bf On Explicit Point Multi-Monopoles in SU(2) Gauge Theory}

\vspace{15mm}

{\Large Alexander D. Popov~$^*$} 
\vskip 1cm
{\em Institut f\"{u}r Theoretische Physik\\
Universit\"{a}t Hannover\\
Appelstra{\ss}e 2, 30167 Hannover, Germany}\\[5mm]
{Email: popov@itp.uni-hannover.de}

\vspace{20mm}

\begin{abstract}
\noindent 
It is well known that the Dirac monopole solution
with the U(1) gauge group embedded into the group SU(2) is
equivalent to the SU(2) Wu-Yang point monopole solution having no
Dirac string singularity. We consider a multi-center configuration 
of $m$ Dirac monopoles and $n$ anti-monopoles and its 
embedding into SU(2) gauge theory. Using geometric methods, 
we construct an explicit solution of the SU(2) Yang-Mills 
equations which generalizes the Wu-Yang solution to the case 
of $m$ monopoles and $n$ anti-monopoles located at arbitrary 
points in $\R^3$.

\end{abstract}

\end{center}

\vfill

\textwidth 6.5truein
\hrule width 5.cm
\vskip.1in

{\small
\noindent ${}^*$
On leave from Bogoliubov Laboratory of Theoretical Physics, JINR,
Dubna, Russia}
\end{titlepage}

\section{Introduction} 
\noindent
Abelian magnetic monopoles play a key role in the dual 
superconductor mechanism of confinement~\cite{H1}
which has been confirmed by many numerical simulations of the
lattice gluodynamics (see e.g.~\cite{Gia, Cher} and references
therein). Due to a dominant role of abelian monopoles
in the confinement phenomena, it is important to understand
better how do they arise in nonabelian pure gauge theories.

A spherically-symmetric monopole solution of the SU(2)
pure gauge field equations was obtained by Wu and Yang
in 1969~\cite{WY1}. This solution is singular at the origin
and smooth on $\R^3-\{0\}$. Initially it was thought that 
it is genuinely nonabelian, yet later it was shown~\cite{WY2}
that this solution is nothing but the abelian Dirac 
monopole~\cite{D} in disguise. Note that the gauge potential
of the finite-energy spherically symmetric 't~Hooft-Polyakov
monopole~\cite{HP} approaches just the Wu-Yang gauge potential
for large $r^2=x^ax^a$.

In this note, we generalize the Wu-Yang solution to a
configuration describing $m$ monopoles and $n$ anti-monopoles
with arbitrary locations in $\R^3$. This explicit solution to the 
Yang-Mills equations can also be used as a guide to the 
asymptotic $r\to\infty$ behaviour of unknown finite-energy
solutions in Yang-Mills-Higgs theory, whose form for small $r$
is determined by multiplying the solution by arbitrary 
functions and minimizing the energy functional, as was proposed 
in~\cite{A}.

\bigskip    

\section{Generic U(1) configurations}
\noindent
We consider the configuration of $m$ Dirac monopoles and $n$
anti-monopoles located at points $\vec a_i=\{a^1_i, a^2_i, a^3_i \}$ 
with $i = 1,\ldots ,m$ and  $i= m+1,\ldots ,m+n$, respectively.
There are delta-function sources for the magnetic field at these points.

Let us introduce the following two regions in $\R^3$:
\begin{equation}\non
R^3_{N,m+n}:=\R^3 - \left\{
\mathop{\bigcup}\limits_{i=1}^{m+n}
(x^1=a^1_i,\, x^2=a^2_i,\, x^3\le a^3_i)\right\}\ ,
\end{equation} 
\begin{equation}\label{RS}
R^3_{S,m+n}:=\R^3 - \left\{
\mathop{\bigcup}\limits_{i=1}^{m+n}
(x^1=a^1_i,\, x^2=a^2_i,\, x^3\ge a^3_i)\right\}\ .
\end{equation} 
{}For simplicity we restrict ourselves to the {\it generic} case
\begin{equation}\label{3}
a^{1,2}_i\ne a^{1,2}_j \for i\ne j\ ,
\end{equation} 
when
\begin{equation}
R^3_{N,m+n}\ \cup\ R^3_{S,m+n}= \R^3- \left\{
\vec a_1,\ldots ,\vec a_{m+n}\right\}\ ,
\end{equation} 
and the two open sets are enough for describing the above 
$(m,n)$-configuration. Namely, the generic configuration of $m$
Dirac monopoles and $n$ anti-monopoles is described by the
gauge potentials
\begin{equation}\label{A}
A^{N,m+n}=\sum\limits_{j=1}^m A^{N,j} + \sum\limits_{j=m+1}^{m+n} 
\bar A^{N,j}
\und
A^{S,m+n}=\sum\limits_{j=1}^m A^{S,j} + \sum\limits_{j=m+1}^{m+n}
 \bar A^{S,j}\ ,
\end{equation} 
where $A^{N,m+n}$ and $A^{S,m+n}$ are well defined on $R^3_{N,m+n}$ 
and $R^3_{S,m+n}$, respectively. 
Here
\begin{equation}\label{ANj}
A^{N,j}=A^{N,j}_a\diff x^a \with 
A^{N,j}_1=\frac{\im x^2_j}{2r_j(r_j+x^3_j)}\ ,\quad
A^{N,j}_2=-\frac{\im x^1_j}{2r_j(r_j+x^3_j)}\ ,\quad
A^{N,j}_3=0\ ,
\end{equation} 
\begin{equation}\label{ASj}
A^{S,j}=A^{S,j}_a\diff x^a \with 
A^{S,j}_1= -\frac{\im x^2_j}{2r_j(r_j-x^3_j)}\ ,\quad
A^{S,j}_2= \frac{\im x^1_j}{2r_j(r_j-x^3_j)}\ ,\quad
A^{S,j}_3=0\ ,
\end{equation} 
where
\begin{equation}
x^c_j=x^c - a^c_j\ ,\quad r^2_j=\de_{ab}x^a_jx^b_j\ ,
\quad a,b,c=1,2,3\ ,
\end{equation} 
and $\bar A^{N,j} = -A^{N,j}$, $\bar A^{S,j} = -A^{S,j}$.
On the intersection $R^3_{N,m+n}\cap R^3_{S,m+n}$ we have
\begin{equation}\label{AA}
A^{N,m+n}=A^{S,m+n}+\diff \, \ln\left (\prod\limits_{i=1}^{m}
\left (\frac{\bar y_i}{y_i}\right )^{\frac{1}{2}}
\prod\limits_{j=m+1}^{m+n}
\left (\frac{y_j}{\bar y_j}\right )^{\frac{1}{2}}\right )\ ,
\end{equation} 
where $y_j=x_j^1+\im x_j^2$ and bar denotes a complex conjugation.

{\bf Remark.} Note that in the case when $a_i^{1,2}=a_j^{1,2}$ 
for some $i\ne j$,
one has to introduce more than two open sets covering the space
$\R^3-\{\vec a_1, \ldots , \vec a_{m+n}\}$ and  define  gauge 
potentials on each of these sets as well as transition functions 
on their intersections. However, for the case 
$\vec a_1= \ldots =\vec a_{m+n}=\vec a$ the two sets
(\ref{RS}) are again enough to cover $\R^3-\{\vec a\}$  and 
the gauge potential (\ref{A})-(\ref{ASj}) will describe $m-n$ 
monopoles (if $m>n$) or $n-m$ anti-monopoles (if $m<n$) sitting 
on top of each other.

One can simplify expressions (\ref{A})-(\ref{AA}) by introducing 
functions of coordinates
\begin{equation}\label{ywv}
w_j := \frac{y_j}{r_j -x_j^3 }=\e^{\im\vp_j}\cot\frac{\th_j}{2}
\und 
v_j := \frac{1}{w_j}=\frac{\bar y_j}{r_j + x_j^3 }=
\e^{-\im\vp_j}\tan\frac{\th_j}{2}\ ,
\end{equation} 
where
\begin{equation}\label{x}
x_j^1=r_j\,\sin\th_j\,\cos\vp_j\ ,\quad
x_j^2=r_j\,\sin\th_j\,\sin\vp_j\und
x_j^3=r_j\,\cos\th_j\ .
\end{equation} 
Note that $w_i\to\infty$ for $x^{1,2}\to a^{1,2}_i,\, 
x^3\ge a^{3}_i$, and $v_i\to\infty$ for $x^{1,2}\to a^{1,2}_i,
\, x^3\le a^{3}_i$. In terms of $w_j$ and $v_j$ the gauge 
potentials (\ref{A})-(\ref{ASj}) have the form
\begin{equation}\label{ANmn}
A^{N,m+n}= \sum\limits_{i=1}^m\frac{1}{2(1+v_i\bar v_i)}
(\bar v_i \diff v_i - v_i\diff\bar v_i )+
\sum\limits_{i=m+1}^{m+n}\frac{1}{2(1+v_i\bar v_i)}
(v_i\diff\bar v_i -\bar v_i \diff v_i )\ ,
\end{equation} 
\begin{equation}\label{ASmn}
A^{S,m+n}=\sum\limits_{i=1}^m\frac{1}{2(1+w_i\bar w_i)}
(\bar w_i \diff w_i - w_i\diff\bar w_i )+
\sum\limits_{i=m+1}^{m+n}\frac{1}{2(1+w_i\bar w_i)}
(w_i\diff\bar w_i -\bar w_i \diff w_i )\ .
\end{equation} 
On the intersection $R^3_{N,m+n}\cap R^3_{S,m+n}$ of two domains
(\ref{RS}) these configurations are related by the transformation
\begin{equation}\label{cap}
A^{N,m+n}=
A^{S,m+n}+\diff \, \ln\!\left (\prod\limits_{i=1}^{m}
\left (\frac{\bar w_i}{w_i}\right )^\sfrac{1}{2}{}
\prod\limits_{j=m+1}^{m+n}
\left (\frac{w_j}{\bar w_j}\right )^{\sfrac{1}{2}}\right ),
\end{equation} 
since ${\bar y_i}/{y_i}={\bar w_i}/{w_i}$. Note that the transition
function in (\ref{cap}) can also be written in terms of $v_i$ by
using the relation ${v_i}/{\bar v_i}={\bar w_i}/{w_i}$.

{}For the abelian curvature $F^{D,m+n}$ we have
\begin{equation}\label{FDmn}
F^{D,m+n}= \diff \,A^{N,m+n}=-\sum\limits_{i=1}^m
\frac{\diff v_i\wedge\diff\bar v_i}{(1+v_i\bar v_i)^2}
  + \sum\limits_{i=m+1}^{m+n}
\frac{\diff v_i\wedge\diff\bar v_i}{(1+v_i\bar v_i)^2}=\non
\end{equation} 
\begin{equation}
=-\sum\limits_{i=1}^m\frac{\diff w_i\wedge\diff\bar w_i}
{(1+w_i\bar w_i)^2}  + \sum\limits_{i=m+1}^{m+n}
\frac{\diff w_i\wedge\diff\bar w_i}{(1+w_i\bar w_i)^2}
=\diff \,A^{S,m+n}\ .
\end{equation} 
It is not difficult to see that $F^{D,m+n}$ is singular only at
points $\{\vec a_1,\ldots ,\vec a_{m+n}\}$, where monopoles and 
anti-monopoles are located.

\bigskip

\section{Point SU(2) configurations}
\noindent
The generalization of the Wu-Yang SU(2) monopole~\cite{WY1} to a 
configuration describing
$m$ monopoles and $n$ anti-monopoles can be obtained as follows.
Let us multiply equation (\ref{cap}) by the Pauli matrix $\s_3$
and rewrite it as 
\begin{equation}\label{AN}
A^{N,m+n}\s_3 = f_{NS}^{(m,n)}A^{S,m+n}\s_3 \bigl(f_{NS}^{(m,n)}
\bigr)^{-1}+f_{NS}^{(m,n)}\diff \bigl(f_{NS}^{(m,n)}\bigr)^{-1}\ ,
\end{equation} 
where 
\begin{equation}\label{f}
f_{NS}^{(m,n)}=\left(\begin{matrix} \prod\limits_{i=1}^{m}
\left (\frac{w_i}{\bar w_i}\right )^{\frac{1}{2}}
\prod\limits_{j=m+1}^{m+n}
\left (\frac{\bar w_j}{w_j}\right )^{\frac{1}{2}}&0 \\0& 
\prod\limits_{i=1}^{m}
\left (\frac{\bar w_i}{w_i}\right )^{\frac{1}{2}}
\prod\limits_{j=m+1}^{m+n}
\left (\frac{w_j}{\bar w_j}\right )^{\frac{1}{2}}
 \end{matrix}\right)\ .
\end{equation} 
It can be checked by direct calculation that  the transition matrix 
(\ref{f}) can be splitted as 
\begin{equation}\label{fNS}
f_{NS}^{(m,n)}=(g_{N}^{(m,n)})^{-1}g_{S}^{(m,n)}\ ,
\end{equation} 
where the $2\times 2$ unitary matrices
\begin{equation}\label{gN}
g_{N}^{(m,n)}= \frac{1}{\bigl(1+ 
\prod\limits_{i=1}^{m+n} v_i \bar v_i
\bigr)^{\frac{1}{2}}}
\begin{pmatrix}\prod\limits_{j=1}^{m}v_j
\prod\limits_{k=m+1}^{m+n}\bar v_k&1\\
-1&\prod\limits_{j=1}^{m} \bar v_j
\prod\limits_{k=m+1}^{m+n}v_k\end{pmatrix}
\end{equation} 
and
\begin{equation}\label{gS}
g_{S}^{(m,n)}= \frac{1}{\bigl(1+ \prod\limits_{i=1}^{m+n} 
w_i\bar w_i\bigr)^{\frac{1}{2}}}
\begin{pmatrix}1&\prod\limits_{j=1}^{m}\bar w_j
\prod\limits_{k=m+1}^{m+n}w_k\\
-\prod\limits_{j=1}^{m} w_j\prod\limits_{k=m+1}^{m+n}
\bar w_k&1  \end{pmatrix}
\end{equation} 
are well defined on $R^3_{N,m+n}$ and $R^3_{S,m+n}$, respectively.
Using formulae (\ref{ywv}) and (\ref{x}), one can rewrite these 
matrices in the coordinates $x^a_i$ with explicit dependence on 
moduli $\vec a_i$ for $i=1,\ldots ,m+n$.

Substituting  (\ref{fNS}) into (\ref{AN}), we obtain 
\begin{equation}\label{ASU2}
A^{N,m+n}g_{N}^{(m,n)}\s_3\bigl(g_{N}^{(m,n)}\bigr)^\+{+}
g_{N}^{(m,n)}\diff \bigl(g_{N}^{(m,n)}\bigr)^\+{=}
 A^{S,m+n}g_{S}^{(m,n)}\s_3\bigl( g_{S}^{(m,n)}\bigr)^\+{+}
g_{S}^{(m,n)}\diff \bigl(g_{S}^{(m,n)}\bigr)^\+{=:}A^{(m,n)}_{su(2)},
\end{equation} 
where by construction $A_{su(2)}^{(m,n)}$ is well defined on 
$R^3_{N,m+n}\cup R^3_{S,m+n}=\R^3-\{\vec a_1,\ldots ,\vec a_{m+n}\}$. 
Geometrically, the existence of splitting (\ref{fNS}) means that 
Dirac's nontrivial U(1) bundle over $\R^3-\{\vec a_1,\ldots ,
\vec a_{m+n}\}$ trivializes when being embedded into an SU(2) bundle. 
The matrices (\ref{gN}) and (\ref{gS}) define this trivialization 
since $f_{NS}^{(m,n)}\mapsto\tilde f_{NS}^{(m,n)}= 
g_{N}^{(m,n)}f_{NS}^{(m,n)}\bigl(g_{S}^{(m,n)}\bigr)^{-1}={\bf 1}_2$.

{\bf Remark.} Recall that we consider generic configurations with the
conditions (\ref{3}). In the case of $a_i^{1,2}$ coinciding for 
some $i\ne j$, one has   $R^3_{N,m+n}\cup R^3_{S,m+n}\ne\R^3-\{\vec a_1,
\ldots , \vec a_{m+n}\}$ and the gauge potential (\ref{ASU2}) can
have singularities outside $R^3_{N,m+n}\cup R^3_{S,m+n}$. For example,
in the case $m=2, n=0$, $a_1^{1,2}=a_2^{1,2}=0$ and $a_1^3=-a_2^3=a$,
the gauge potential describing two separated monopoles will be
singular on the interval $-a\le x^3\le a$. To have nonsingular
$A_{su(2)}^{(2,0)}$ one should consider $a_1^{1,2}\ne a_2^{1,2}$ or to
use three open sets covering $\R^3-\{\vec a_1,\vec a_{2}\}$ instead of 
two ones.

The field strength for the configuration (\ref{ASU2}) is given by
\begin{equation}\label{FSU2}
F^{(m,n)}_{su(2)}=\diff A^{(m,n)}_{su(2)}+A^{(m,n)}_{su(2)}\wedge 
A^{(m,n)}_{su(2)}= \im  F^{D,m+n} Q_{(m,n)}\ ,
\end{equation} 
where the su(2)-valued matrix
\begin{equation}\label{Qmn}
Q_{(m,n)}:=-\im g_{N}^{(m,n)}\s_3 (g_{N}^{(m,n)})^{\+}=
-\im g_{S}^{(m,n)}\s_3 (g_{S}^{(m,n)})^{\+}
\end{equation} 
is well defined on $R^3_{N,m+n}\cup R^3_{S,m+n}$. It is easy to see that
$Q_{(m,n)}^2=-1$ and $Q_{(m,n)}$ may be considered as the generator 
of the group U(1) embedded into SU(2). Then the abelian nature of the 
configuration (\ref{ASU2})-(\ref{FSU2}) becomes obvious. Furthermore, 
for 
\begin{equation}\label{AF}
A^{(m,n)}_{su(2)}=A^{(m,n)}_a\diff x^a \und 
F^{(m,n)}_{su(2)}=\frac{1}{2}F^{(m,n)}_{ab}\diff x^a \wedge \diff x^b
\end{equation} 
one can easily show that
\begin{equation}
\pa_a F^{(m,n)}_{ab} + [A^{(m,n)}_a,\, F^{(m,n)}_{ab} ]=
\im\, \bigl(\pa_a F^{D,m+n}_{ab} \bigr)Q_{(m,n)}
\end{equation} 
and therefore on the space $\R^3-\{\vec a_1,\ldots ,\vec a_{m+n}\}$
we have 
\begin{equation}
\pa_a F^{(m,n)}_{ab} + [A^{(m,n)}_a,\, F^{(m,n)}_{ab} ]=0\ ,
\end{equation} 
which follows from the field equations describing $m$ Dirac 
monopoles and $n$ anti-monopoles.
Note that the solution (\ref{ASU2})-(\ref{AF}) of the SU(2)
gauge theory can be embedded in any larger gauge theory following 
e.g.~\cite{BW}. 

\bigskip  

\section{Point monopoles via  Riemann-Hilbert problems}
\noindent
Here we want to rederive the described configurations by solving
a matrix Riemann-Hilbert problem. For simplicity, we restrict 
ourselves to the case of $m$ monopoles.
 
Let us consider the Bogomolny equations~\cite{Bo}
\begin{equation}\label{Bo}
F_{ab}=\eps_{abc}\,D_c\c\ ,
\end{equation} 
where $D_c=\pa_c+[A_c,\,\cdot\ ]$ and the fields $A_a, F_{ab}=
\pa_aA_b-\pa_bA_a+[A_a,A_b ]$ and $\c$ take values in the Lie 
algebra $u(q)$. Obviously, in the abelian case $D_c\c = \pa_c\c$. 
Note that for the gauge 
fields $F^{D,m}$ given by 
(\ref{FDmn}) we have
\begin{equation}\label{FDm}
F_{ab}^{D,m}=\eps_{abc}\,\pa_c\p^{(m)}\with 
\p^{(m)}=\sum\limits_{k=1}^{m}\frac{\im}{2r_k}\ .
\end{equation} 
Analogously, for the field $F^{(m)}_{ab}$ from (\ref{AF})
we have
\begin{equation}\label{Fabm}
F_{ab}^{(m)}=\eps_{abc}\,D_c\Phi^{(m)}\with
\Phi^{(m)}= \im\, \phi^{(m)} Q_{(m)}\ ,
\end{equation} 
where $\phi^{(m)}$ is given in (\ref{FDm}) and $Q_{(m)}$ in 
(\ref{Qmn}). Thus, both U(1) and SU(2) multi-monopoles as well as
$(m,n)$-configurations (\ref{ANmn})-(\ref{FDmn}) and 
(\ref{ASU2})-(\ref{Qmn})
can be considered as solutions of the Bogomolny equations (\ref{Bo}).
In fact, the second order pure Yang-Mills equations for $F^{D,m}_{ab}$
and $F^{(m)}_{ab}$ can be obtained by differentiating (\ref{FDm}) and
(\ref{Fabm}), respectively. Moreover, in pure SU(2) Yang-Mills theory
in (3+1)-dimensional Minkowski space-time, one can choose the component 
$A_0$ of the gauge potential $A{=}A_0\diff t{+}A_a\diff x^a$ to be nonzero
and proportional to $\Phi^{(m)}$ (the abelian case is similar). Then the
configuration $\{A_0^{(m)}, A_a^{(m)} \}$ will be a static multi-dyon 
solution of the Yang-Mills equations.

Recall that the Bogomolny equations (\ref{Bo}) can be obtained 
as the compatibility conditions of the linear system
\begin{equation}\label{linsys}
\bigl [D_{\bar y}-\frac{\la}{2}(D_3+\im\c )\bigr ]\j =0\und
\bigl [\frac{1}{2}(D_3-\im\c )+\la D_y\bigr ]\j =0\ ,
\end{equation} 
where $D_{\bar y}=\frac{1}{2}(D_1+\im D_2),\ D_{y}=\frac{1}{2}(D_1-
\im D_2)$ and the auxiliary $q\times q$ matrix $\j (x^a, \la )$ depends 
holomorphically on a new variable $\la\in U\subset\C P^1$. Such matrices 
$\psi$ can be found via solving a parametric Riemann-Hilbert problem 
which is formulated in the monopole case as follows~\cite{WF}. Suppose 
we are given a $q\times q$ matrix $f_{+-}$ depending holomorphically on 
\begin{equation}\label{eta}
\h =y-2\la x^3-\la^2\bar y
\end{equation} 
and $\la$
for $\la\in U_+\cap U_-$, where $U_+=\C P^1 -\{\infty\}$ and 
$U_-=\C P^1 -\{0\}$.
Then for each fixed $(x^a)\in\R^3$ and $\la\in S^1\subset U_+\cap U_-$ 
one should factorize this matrix-valued function,
\begin{equation}\label{f+-}
f_{+-}(x, \la) = \j^{-1}_+(x, \la)\j_-(x, \la)\ ,
\end{equation} 
in such a way that $\j_+$ and $\j_-$ extend  holomorphically in $\la$ 
onto subsets of $U_+$ and $U_-$, respectively. In order to insure 
that $A_a^\+=-A_a$ and $\c^\+ = -\c$ in (\ref{linsys}) with 
$\psi = \psi_{\pm}$ one should also impose the (reality) conditions
\begin{equation}\label{Cond1}
f_{+-}^\+ (x, -{\bar\la}^{-1})\ =\ f_{+-} (x,\la)\und
\psi_+^\+(x, - {\bar\la}^{-1})\ =\ \psi_-^{-1}(x, {\la})
\ .
\end{equation} 
After finding such $\psi_{\pm}$ for an educated guess of $f_{+-}$,
one can get $A_a$ and $\c$ from the linear system (\ref{linsys})
with the matrix function $\psi_+$ or $\psi_-$ instead of $\psi$.
Namely, from (\ref{linsys}) we get
\begin{equation}\label{AAAA1}
A_{\bar y}:=\sfrac{1}{2}(A_1+\im A_2)\ =\ 
\psi_+\pa_{\bar y}\psi_+^{-1}|_{\la=0}\ ,\qquad
A_3-\im\c\ =\ 
\psi_+\pa_{3}\psi_+^{-1}|_{\la=0}\ ,
\end{equation} 
\begin{equation}\label{AAAA2}
A_y:=\sfrac{1}{2}(A_1 - \im A_2)\ =\ 
\psi_-\pa_y\psi_-^{-1}|_{\la =\infty}\ ,\qquad
A_3 +\im \c \ =\ 
\psi_-\pa_3\psi_-^{-1}|_{\la =\infty}\ .
\end{equation}
{}For more details see~\cite{WF,LP} and references therein.

The construction of U(1) multi-monopole solutions via solving the
Riemann-Hilbert problem for the function
\begin{equation}\label{trfun2} 
f_{+-}^{D,m} = \frac{\la^m}{\prod_{k=1}^m {\h_k}}=:\rho_m
\with \h_k = \h - h(a^1_k,a^2_k,a^3_k,\la )=
(1{-}\la^2)x^1_k+\im(1{+}\la^2)x^2_k-2\la x^3_k
\end{equation} 
was discussed in~\cite{LP} and here we describe only the SU(2) case. 
The ansatz for $f_{+-}^{(m)}$ which satifies (\ref{Cond1}) 
only for odd $m$ was written down in the appendix C of~\cite{LP}. 
Here we introduce the ansatz 
\begin{equation}\label{tildef}
f_{+-}^{(m)}=
\begin{pmatrix}\rho_{m}&\la^{-m}\\
(-1)^m\la^{m}&\rho^{-1}_{m}+(-1)^m\rho^{-1}_{m}\end{pmatrix}  
\end{equation} 
satisfying the reality condition (\ref{Cond1}) for any $m$. It is not 
difficult to see that 
\begin{equation}\label{tldf}
f_{+-}^{(m)}=\begin{pmatrix}1&0\\(-1)^m\la^m\rho_m^{-1}&1\end{pmatrix}
\begin{pmatrix}f_{+-}^{D,m}&0\\0&(f_{+-}^{D,m})^{-1}\end{pmatrix}
\begin{pmatrix}1&\la^{-m}\rho_m^{-1}\\0&1\end{pmatrix}\sim  
\begin{pmatrix}f_{+-}^{D,m}&0\\0&(f_{+-}^{D,m})^{-1}\end{pmatrix}\ ,
\end{equation} 
where the diagonal matrix in (\ref{tldf}) describes the Dirac
line bundle $L$ (the U(1) gauge group) embedded into the rank 2
complex vector bundle (the SU(2) gauge group) as $L\oplus L^{-1}$.
This gives another proof of the equivalence of U(1) and SU(2) point
monopole configurations (see~\cite{LP} for more details). 
Furthermore, the matrix (\ref{tildef}) can be splitted as follows:
\begin{equation}\label{split}
f_{+-}^{(m)}= \bigr(\j_+^{(m)}\bigl)^{-1} \j_-^{(m)}\ ,
\end{equation} 
where
\begin{equation}\label{tldj}
\j_+^{(m)}=\hat\j_+^{(m)}
\begin{pmatrix}1&0\\(-1)^{m+1}\la^m\rho_m^{-1}&1\end{pmatrix}
\ , \quad
\j_-^{(m)}=\hat\j_-^{(m)}
\begin{pmatrix}1&-\la^{-m}\rho_m^{-1}\\0&1\end{pmatrix}\ ,
\end{equation} 
\begin{equation}\label{+-j}
\hat\j_+^{(m)}=g_S^{(m)}
\begin{pmatrix}\j_+^{S,m}&0\\0&(\j_+^{S,m})^{-1}\end{pmatrix}
\ , \quad
\hat\j_-^{(m)}=g_N^{(m)}
\begin{pmatrix}\j_-^{N,m}&0\\0&(\j_-^{N,m})^{-1}\end{pmatrix}\ ,
\end{equation} 
\begin{equation}\label{j+}
\j_+^{S,m}= \prod\limits_{i=1}^{m}\j_+^{S}(x_i^a,\la )\ ,\quad 
\j_+^{S}(x_i^a,\la )=\xi_+(x_i^a )-\la \xi_+^{-1}(x_i^a )\bar y_i
\ ,\quad  \xi_+(x_i^a )=(r_i-x_i^3)^\frac{1}{2}\ ,
\end{equation} 
\begin{equation}\label{j-}
\j_-^{N,m}= \prod\limits_{i=1}^{m}\j_-^{N}(x_i^a,\la )\ ,\quad 
\bigr(\j_-^{N}(x_i^a,\la )\bigl)^{-1}=
\xi_-(x_i^a )\bar y_i +\la^{-1} \xi_-^{-1}(x_i^a )\ ,\quad  
\xi_-(x_i^a )=(r_i+x_i^3)^{-\frac{1}{2}}\ .
\end{equation} 
The explicit form of $g_N^{(m)}$ and $g_S^{(m)}$ is given in 
(\ref{gN}) and (\ref{gS}). Note that both $\j_\pm^{(m)}$ and
$\hat\j_\pm^{(m)}$ satisfy the reality conditions (\ref{Cond1}).

Formulae (\ref{split})-(\ref{j-}) solve the parametric 
Riemann-Hilbert problem for our $f_{+-}^{(m)}$ restricted to a 
contour on $\C P^1$ which avoids all zeros of the function 
$\prod\limits_{k=1}^{m}\h_k$. Substituting (\ref{tldj})-(\ref{j-}) 
into formulae (\ref{AAAA1})-(\ref{AAAA2}), we get
\begin{equation}\label{AAA}
A_{\bar y}^{(m)}=\hat g_S^{(m)}\pa_{\bar y}
\bigr(\hat g_S^{(m)}\bigl)^{-1},
\
A_{y}^{(m)}=\hat g_N^{(m)}\pa_{y}\bigr(\hat g_N^{(m)}\bigl)^{-1},
\
A_{3}^{(m)}=g_S^{(m)}\pa_{3}\bigr(g_S^{(m)}\bigl)^{\+}
=g_N^{(m)}\pa_{3}\bigr(g_N^{(m)}\bigl)^{\+},
\end{equation} 
\begin{equation}\label{chi}
\c^{(m)}=\frac{\im}{2}\Bigr(\hat g_S^{(m)}\pa_{3}
\bigr(\hat g_S^{(m)}\bigl)^{-1} - \hat g_N^{(m)}\pa_{3}
\bigr(\hat g_N^{(m)}\bigl)^{-1}\Bigl)\ ,
\end{equation}  
where 
\begin{eqnarray}
\hat g_S^{(m)}
%{=}\j^{(m)}_+|_{\la =0}
{=} g_S^{(m)}
\begin{pmatrix}\xi_+&0\\
0&\xi_+^{-1}\end{pmatrix}
\ \mbox{with}&  
g_S^{(m)}{=}\frac{1}{\bigr(\prod\limits_{i=1}^{m}(r_i{-}x_i^3)^2+
\prod\limits_{i=1}^{m}y_i\bar y_i\bigl)^\frac{1}{2} }
\begin{pmatrix}\prod\limits_{j=1}^{m}(r_j{-}x_j^3)&
\prod\limits_{j=1}^{m}\bar y_j\\{-}\prod\limits_{j=1}^{m} y_j
&\prod\limits_{j=1}^{m}(r_j{-}x_j^3)\end{pmatrix},&
\label{hatgS}
\\
\hat g_N^{(m)}
%{=}\j^{(m)}_-|_{\la =\infty}
{=}g_N^{(m)}
\begin{pmatrix}\xi_-^{-1}&0\\
0&\xi_-\end{pmatrix}
\ \mbox{with}& 
g_N^{(m)}{=}\frac{1}{\bigr(\prod\limits_{i=1}^{m}(r_i+x_i^3)^2+
\prod\limits_{i=1}^{m}y_i\bar y_i\bigl)^\frac{1}{2} }
\begin{pmatrix}\prod\limits_{j=1}^{m}\bar y_j&
\prod\limits_{j=1}^{m}(r_j{+}x_j^3)\\
{-}\prod\limits_{j=1}^{m}(r_j{+}x_j^3)&
\prod\limits_{j=1}^{m} y_j\end{pmatrix}&\label{hatgN}
\end{eqnarray}
and 
\begin{equation}
\xi_+=\prod\limits_{k=1}^{m}\xi_+(x_k^a )=
\prod\limits_{k=1}^{m}(r_k{-}x_k^3)^\sfrac{1}{2}
\ ,\quad 
\xi_-=\prod\limits_{k=1}^{m}\xi_-(x_k^a )=
\prod\limits_{k=1}^{m}(r_k{+}x_k^3)^{-\sfrac{1}{2}}\ .
\end{equation}
It is not difficult to see that the configuration (\ref{AAA})
coincides with (\ref{ASU2}) and $\c^{(m)}$ from (\ref{chi}) 
with $\Phi^{(m)}$ from (\ref{Fabm}).
Thus, we have derived SU(2) multi-monopole point-like solutions
via a parametric Riemann-Hilbert problem.

\bigskip

\noindent
{\bf Acknowledgements}

\medskip

\noindent
I thank O.~Lechtenfeld and H.~M\"uller-Ebhardt for useful comments.
This work was supported by the Deutsche Forschungsgemeinschaft 
(DFG).

\end{document}